\documentstyle[]{elsart}
\input{psfig.sty}
\begin{document}
\runauthor{G. Ghisellini}
\begin{frontmatter}
\title{Extreme blazars}
\author[]{Gabriele Ghisellini}
\address[]{Osservatorio Astronomico di Brera, Merate, Italy}

\begin{abstract}
The recent Cherenkov telescope observations and detections of the 
BL Lac objects Mkn 421, Mkn 501, 1ES 2344+514, PKS 2155--304 and possibly
1ES 1959+658 have shown that there exists a subclass of BL Lac
objects emitting a substantial fraction of their power between 
the GeV and the TeV bands.
These are the sources whose synchrotron spectrum peaks, in a 
$\nu$--$\nu F_{\nu}$ representation, in the EUV or X--ray band.
Here I suggest that even more extreme BL Lacs can exist,
whose synchrotron spectrum peaks in the MeV band.
These sources should emit a substantial fraction of their power in the 
TeV band by the inverse Compton process.
Limits to the maximum possible emitted frequencies are discussed.
\end{abstract}


\end{frontmatter}

\section{Introduction}
Although united in a single class because of their similar properties, 
blazars come in different flavors, according to the strength of 
their emission lines, the level of their optical polarization 
and the ratio between their X--ray and radio fluxes.
The most important recent realization has been the 
discovery that blazars are powerful $\gamma$--ray emitters: at least 
during flares, they can emit in this band up to 90\% of their 
bolometric output 
(von Montigny et al. 1995; Thompson et al. 1995, 1996;
Weekes et al. 1996; Petry et al. 1996).
Thanks to EGRET, onboard CGRO, and to the impressive improvements of
ground based Cherenkov telescopes, we at last know where most of
the blazar power is emitted.
Rapid variability indicates that the $\gamma$--ray and X--ray emitting regions
are compact, and this is yet another proof that the bulk of blazar emission 
is beamed, since otherwise the source
would be opaque to $\gamma$--rays (see e.g. Dondi \& Ghisellini 1995).

On the theoretical side, this discovery led to major improvements in our
understandings of blazars physics in particular and how jet must work
in general.
One important outcome of these studies is the realization that
blazars probably form a sequence in their properties.
As will be explained in somewhat more detailed below,
there is a link between the maximum energy of the emitting
electrons and the intrinsic luminosity of the source.
Therefore there is a link between the overall spectral energy
distribution (SED) and the bolometric luminosity of blazars.
Very high energy electrons are present only in low luminosity sources, 
and these are the ones emitting significantly in the GeV--TeV band.
Can even more extreme objects exist?

\section{The SED of blazars}

We now know that the SED of blazars is characterized by two broad
peaks, the first in the IR--EUV and sometimes X--ray band, and the
second in the MeV--GeV bands.
There is almost unanimity about the interpretation of the first peak
as due to incoherent synchrotron emission.
The only remaining doubt concerns the interpretation of the fast (intraday) 
radio variability observed in a large fraction (about 1/4)  of blazars 
(Wagner \& Witzel 1995).
If it is indeed an intrinsic property, and it is not due to
interstellar scintillation, then this phenomenon will deeply change
the present theoretical ``paradigm", requiring some coherent process
to be at work (Benford \& Lesh 1998).
The origin of the second peak is still uncertain. 
It can be due to synchrotron Self Compton (SSC: Maraschi, Ghisellini \& Celotti 1992;
Bloom \& Marscher 1996) or a mixture of SSC plus a 
contribution by inverse Compton scattering off photons produced externally 
to the jet (EC: Dermer \& Schlickeiser 1993; Sikora, Begelman \& Rees 1994; 
Blandford \& Levinson 1995, Ghisellini \& Madau 1996), 
or another more energetic synchrotron component 
(as in the `proton blazar' model by Mannheim 1993).

In this framework, objects presenting extreme properties are the most
interesting, since they best constrain our models.
In this respect the discovery made by the existing Cherenkov telescopes of 
TeV emission from blazars is extremely important, and stimulates new and 
interesting ideas on the physics of relativistic jets.
If, as the correlated variability suggests, the TeV flux is produced
by the same electrons dominating by synchrotron emission
in the X--ray band, then we have a powerful tool to study 
in detail the acceleration process at its limit.

Even before the advent of EGRET blazars were divided in subclasses
according to their (low energy) SED.
For instance, BL Lacs discovered through radio and X--ray 
surveys were recognized to have different radio to X--ray spectra, 
even if sharing other properties such as the absence of strong 
emission lines, the rapid and large amplitude variability and the 
same average X--ray luminosity.
This led Maraschi et al. (1986) and Ghisellini \& Maraschi (1989) to
try to unify these two blazar subclasses assuming that they simply
corresponded to a different viewing angle under which we see an
accelerating, inhomogeneous jet.

But Giommi \& Padovani (1994) later noticed that the SED of radio and 
X--ray selected BL Lacs showed peaks at different energies, 
and proposed that this difference was intrinsic, and not due to 
orientation effects.
They then divided BL Lac objects into  HBL (high energy peak BL Lac)
and LBL (low energy peak BL Lac), the former being sources preferentially
selected through X--ray surveys, and the latter through radio surveys.
Now we can extend the Giommi \& Padovani's idea
of a moving peak also to the high energy part of the SED, since
we now know that the two peak energies correlate.

\section{The blazar sequence}

We attempted to find regularities in the SED of blazars
by first considering the observational properties of complete
samples of sources, and then by modeling the SED of those
blazars detected in the $\gamma$--ray band, to find their
intrinsic physical quantities.
\begin{figure}
\vskip -7 truecm
\psfig{file=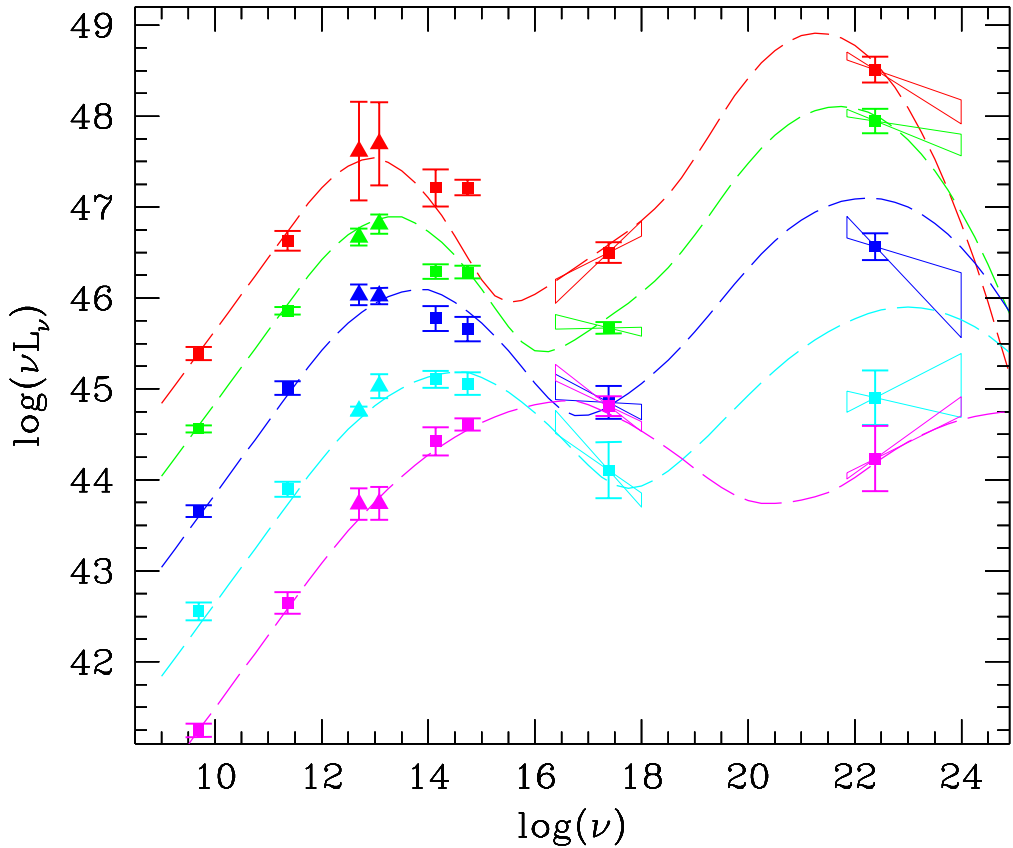} 
\vskip -1 true cm
\caption{The average spectra of blazars,
according to their bolometric luminosity, assumed to be traced by the 
radio luminosity. 
Blazars belonging to complete samples have been
divided in 5 luminosity bins, irrespective of their classification.
Notice how the average SED changes as the overall power changes.
Dashed lines correspond to an analytical, phenomenological fit to the data.
From Fossati et al. 1998.
}
\end{figure}

The first attempt was performed by Fossati et al. (1998), 
considering the SLEW survey of BL Lacs, the 1 Jy sample
of BL Lacs and the 2 Jy sample of flat spectrum radio sources,
for a total of 126 objects.
By collecting data from the literature in selected bands
we could construct the SED of these blazars, to look for trends
with their bolometric luminosities and/or their classification.
To this end we divided all sources in radio luminosity bins (it is 
thought that the radio power traces the bolometric one, see the 
discussion in Fossati et al. 1998), averaging the data of the sources
belonging to the same luminosity bin.
The result is shown in Fig. 1: less powerful objects have the 
synchrotron peak in the soft-medium X--ray range, while their 
high energy peak is at the highest $\gamma$--ray energies.
As the total power increases, both peaks shift to lower frequencies, 
and at the same time the $\gamma$--ray luminosity increases 
its relative importance, arriving to dominate the bolometric output.

\begin{figure}
\vskip -1 truecm
\psfig{file=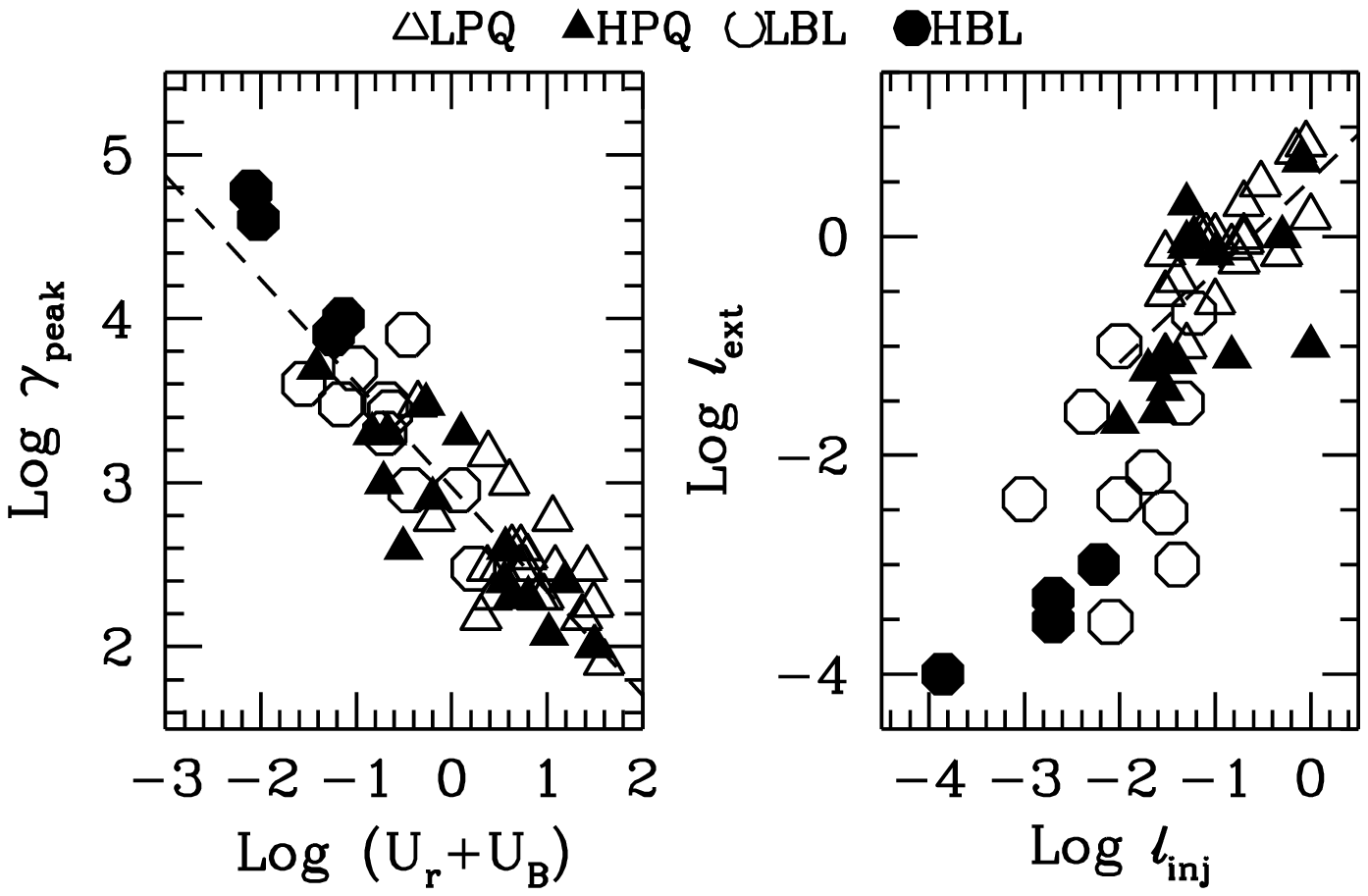} 
\vskip -9.5 true cm
\caption{Correlations found when modeling the SED of
EGRET blazars with an homogeneous EC model.
The left panel shows the correlation between
$\gamma_{peak}$ and the total (magnetic plus radiative, including
the contribution from external photons) energy density.
Different symbols indicate different classes of blazars, as labeled.
The right hand panel shows the correlation between the compactness
$[\ell\equiv L\sigma_{\rm T}/(Rm_{\rm e}c^3)]$ of the external photons and the 
one corresponding to the injected power, as measured in the comoving
frame of the emitting region. From Ghisellini et al. 1998. }
\end{figure}

The approach to fit all blazars with some spectral information in the 
$\gamma$--ray band was undertaken by Ghisellini et al. (1998), using
an homogeneous EC model (but including the SSC contribution).
The main result of this study is the finding of a tight correlation
between the Lorentz factor of the electrons emitting at the peaks, 
$\gamma_{\rm peak}$, and the amount of energy density (both magnetic 
and radiative) present in the emitting region (see Fig. 2).
This in turn correlates with the observed (beamed) luminosity, and the ratio
between the power of the Compton and the synchrotron components.
What is also remarkable is that different subclasses of blazars occupy 
different regions of Fig. 2, indicating a well defined blazar sequence:
\begin{itemize}
\item Low luminosity lineless BL Lacs (HBL) have
large values of $\gamma_{\rm peak}$, synchrotron peak energy in the 
EUV--soft X--rays and a roughly equally powerful Compton
component peaking in the GeV--TeV band.
\item LBL are characterized by a greater overall luminosity, a smaller
$\gamma_{\rm peak}$ and peak energies in the optical and GeV band.
\item More powerful sources, such as HPQ and LPQ, have the smallest values
of $\gamma_{\rm peak}$, peak energies in the mm--far IR and MeV band,
a dominating Compton component in which photons produced externally to the 
jet are more important than the locally produced synchrotron photons.
\item There is a correlation between the intrinsic power and the amount
of external photons needed to fit the spectra. 
As shown in Fig. 2, HBL do not need this external photon component,
while some LBL do (as BL Lac itself, see also Sambruna et al. 1999).
\end{itemize}
One can ask what is the physical reason of the found correlations
between $\gamma_{\rm peak}$ and the energy density in the comoving frame of the
source, the intrinsic power of the source, and the Compton dominance
(i.e. the ratio between the $\gamma$--ray and the IR--optical luminosity).
One possibility is that there is a competition between the acceleration
and the cooling mechanisms in these sources, balancing at $\gamma_{\rm peak}$.
Since we find $\gamma_{\rm peak}\propto U^{-0.6}$, where $U$ is
the sum of the magnetic and radiation energy densities, we
have that the synchrotron and inverse Compton cooling rate
($\propto \gamma^2U$) is, at $\gamma_{\rm peak}$, nearly 
the same for all sources.
This then suggests the presence of some {\it universal} acceleration
mechanism, independent of $\gamma$ and $U$: 
in powerful sources having a large radiation energy density
the balance between gain and losses happens at a small value of 
$\gamma_{\rm peak}$, while in weaker sources this balance is reached
at a larger $\gamma_{\rm peak}$.
A problem with this interpretation is the behavior of 
individual sources, which is sometimes just the
opposite: in the 1997 flare of Mkn 501 the bolometric power increased 
by a factor 20 with respect to the quiescent state,
{\it and $\gamma_{\rm peak}$ increased also, by at least a factor $\sim$10.}
If the mentioned interpretation is correct, one must assume
that during the flare the acceleration rate was much faster
than during quiescence, and this contrasts with the supposed
``universality" of such mechanism.
Further work to solve this discrepancy is needed.

\begin{figure}
\vskip -1 truecm
\psfig{file=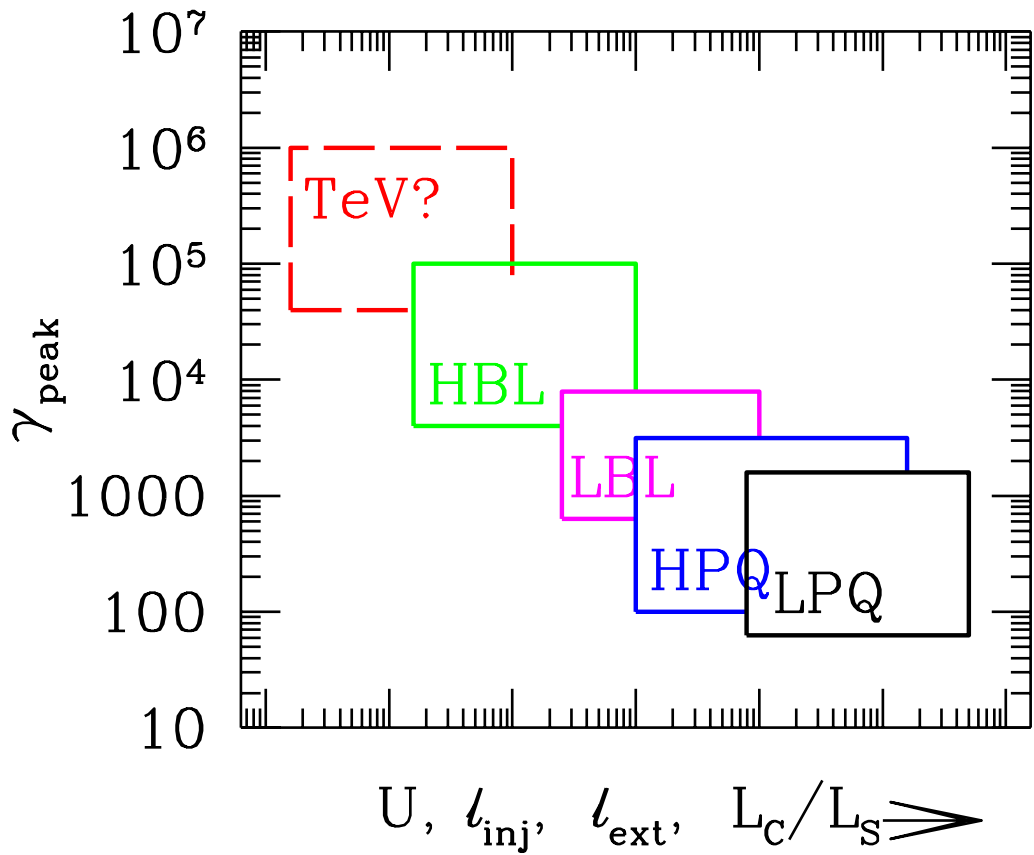} 
\vskip -7 truecm
\caption{
Schematic diagram illustrating the blazar sequence.
The SED of blazars depends on $\gamma_{\rm peak}$, which is
the Lorentz factor of the electrons emitting at the
synchrotron and inverse Compton peaks of the SED.
This correlates with the total energy density in the comoving
frame of the source, its intrinsic power, the amount of
external photons, and the Compton dominance, i.e. the ratio
of the Compton to synchrotron luminosities.
Adapted from Ghisellini et al. 1998.}
\end{figure}

\begin{figure}
\psfig{file=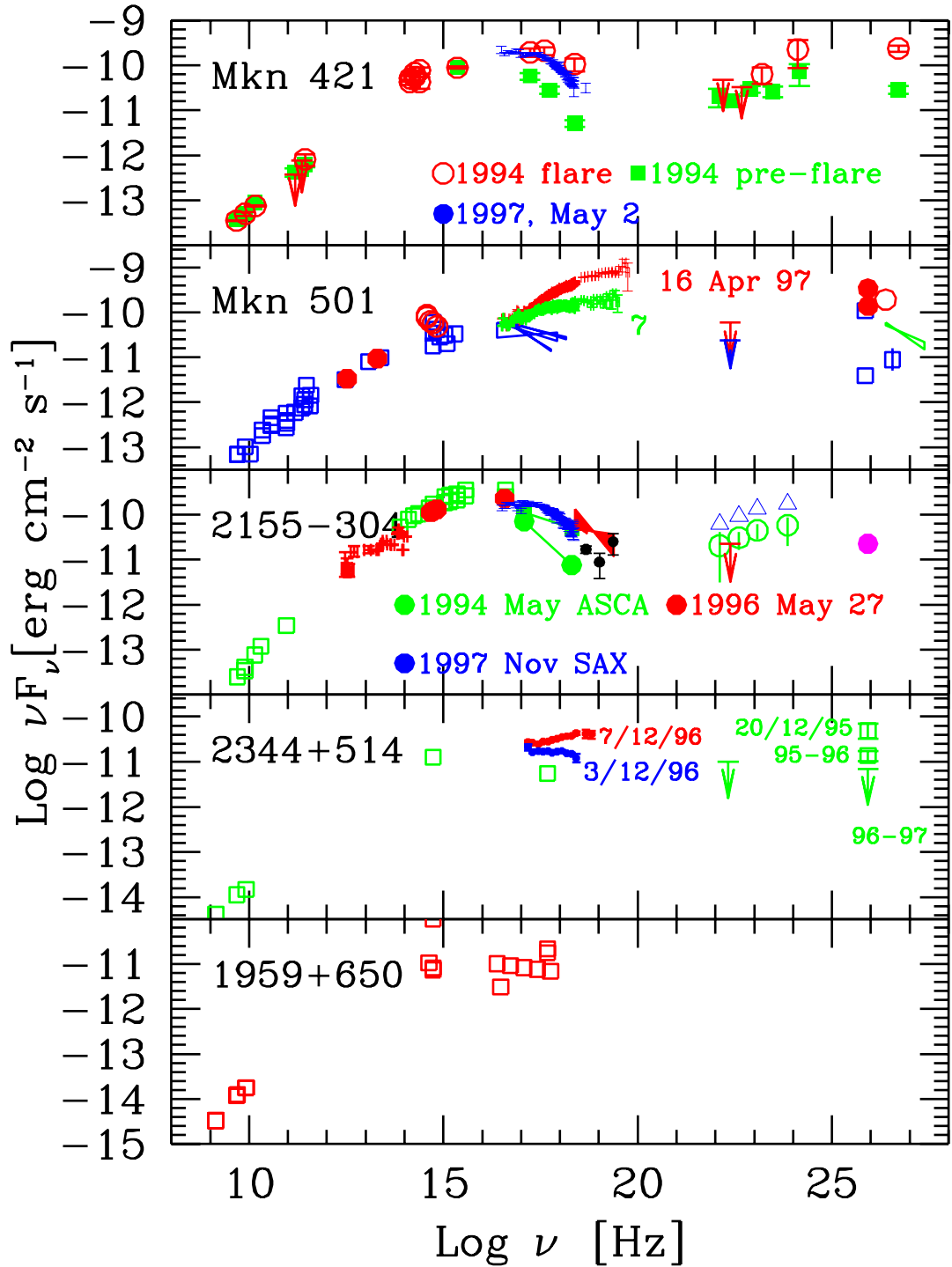} 
\vskip -1 truecm
\caption{The SED of the 5 BL Lac objects already
detected in the TeV band.
Sources of data: 
Mkn 421: Macomb et al. 1995,  Fossati et al. 1998;
Mkn 501: Pian et al. 1998 and references therein;
1ES 2344+514: Giommi, Padovani \& Perlman 1998, Catanese et al. 1997;
PKS 2155--304: Bertone et al., in preparation,
Chiappetti et al. 1998, 
Chadwick et al. 1998;
1ES 1959+658: Perlman et al. 1996.
For this latter source the X--ray slope has been very crudely derived
by the hardness ratio, using archive ROSAT data.}
\end{figure}

\section{TeV BL Lacs}
At this meeting it was announced that 1ES 1959+658 was detected 
in the TeV band, bringing the total number of TeV BL Lacs to 5
(the other four are 
Mkn 421, Punch et al. 1992;
Mkn 501, Quinn et al. 1996; 
1ES 2344+514, Catanese et al. 1997 and 
PKS 2155--304, Chadwick et al. 1998).
The SED of these 5 sources are shown in Fig. 4.
As can be seen their SED is very similar, but note that
Mkn 421 and PKS 2155--304 have always shown a $steep$ ($\alpha_x \sim 1.5$)
X--ray spectrum, while the X--ray slopes of 
Mkn 501 and 1ES 2344+514, during flares, are flat ($\alpha_x < 1$),
making the synchrotron emission to peak in the very hard X--ray range
(Pian et al. 1998; Giommi, Padovani \& Perlman, 1998).

\section{More extreme BL Lacs}

It takes a little leap of imagination to suggest that the
blazar sequence extends somewhat more than what already
discovered by the recent high energy observations.
Here I propose that a new subclass of BL Lac objects can exist,
with lower luminosity and larger $\gamma_{peak}$ than those HBL
already detected at high energies.
To explore this possibility, we must first ask:
\begin{itemize}
\item Why have we not yet detected these extreme objects?
Alternatively: is it possible that we already detected them in some bands, 
but we do not yet know their extreme nature because of
lack of information, such as the X--ray slope and/or the $\gamma$--ray flux?
\item What limits the maximum electron energies?
\end{itemize}

\subsection{Limits on the maximum electron energy: shocks}

In strong shocks, electrons can be accelerated efficiently up to
the energy where their gains equals their energy losses,
which in our case are mainly due to radiation losses.
Assuming, as Guilbert, Fabian \& Rees (1983), that at
each gyroradius the electrons increase their energy by a factor $\gamma$,
and that the radiative losses are dominated by the synchrotron process,
it can be derived that the maximum obtainable $\gamma\propto B^{-1/2}$,
where $B$ is the magnetic field.
This in turn translates in a maximum synchrotron frequency of $\sim$70 MeV,
independent of the magnetic field.
This frequency is then blueshifted by the Doppler factor $\delta$,
yielding an observable maximum frequency of $700(\delta/10)$ MeV.
These values are within the EGRET band, but EGRET did not detect any 
extraordinary blazar whose synchrotron spectrum extends to such
large energies. 
It is therefore unlikely that BL Lacs have 
synchrotron spectra reaching energies greater than $\sim$100 MeV.
There must then be a more severe limit to the maximum observable
synchrotron frequency.

\subsection{Global energetics}

Indeed, another limit can be obtained from the global energetics of jets.
The argument is as follows:
From the observed extended emission of radio sources we know
that at least $10^{45}$--$10^{47}$ erg s$^{-1}$ must be transported
from the black hole to hundreds of kpc, in the form
of Poynting flux and/or bulk kinetic energy of the particles
(e.g. Rawlings \& Saunders 1991).
A fraction of this power can be dissipated into radiation along the way,
and one of the main result of EGRET is to have 
demonstrated that the larger dissipation occurs in a well defined part
of the jet (see Ghisellini and Madau 1996 for a more detailed discussion).
In steady state, the power dissipated into radiation cannot exceed
the power transported by the jet.
Since the synchrotron and inverse Compton losses are $\propto \gamma^2$, 
there is an upper limit to $<\gamma^2>$, corresponding to complete
dissipation.

To be more quantitative, let us define $L_{\rm k}$ and  $L_{\rm B}$ as
the power of bulk kinetic motion of the emitting plasma and of
Poynting flux, respectively
(see e.g. Celotti \& Fabian 1993, Ghisellini \& Celotti 1998):

\begin{equation}
L_{\rm k}\, =\, \pi R^2 \Gamma^2 \beta c \, n^\prime\, m_{\rm e} c^2
(<\gamma>+ m_{\rm p}/m_{\rm e})
\end{equation}
\begin{equation}
L_{\rm B}\, =\, {1 \over 8} R^2 \Gamma^2 \beta c B^2, 
\end{equation}
where $R$ is the cross sectional radius of the jet,
$n^\prime=n/\Gamma$ is the comoving particle density of average energy
$<\gamma>m_{\rm e}c^2$, and $m_{\rm p}$, $m_{\rm e}$ 
are the proton and electron rest masses, respectively. 
An electron proton plasma is assumed.
The synchrotron intrinsic power is 

\begin{equation}
L_{\rm s}^\prime = Volume\,
\int n^\prime (\gamma)\dot \gamma_{\rm s}m_{\rm e}c^2\, d\gamma\, =\,
{2\over 9}\, R^3 \sigma_{\rm T} c n^{\prime} B^2 <\gamma^2>
\end{equation}
where
$\dot\gamma_{\rm s}$ is the synchrotron cooling rate, and 
$n^\prime (\gamma)\propto \gamma^{-p}$ between
$\gamma_{\rm min}$ and $\gamma_{\rm peak}$.
We assume that all electrons partecipating in the
bulk flow are accelerated to random energies $\gamma m_ec^2$
(i.e. the density $n^\prime$ in eq. 1 and 3 is the same).
For a viewing angle $\sim 1/\Gamma$, the luminosity calculated assuming 
isotropy is related to $L_{\rm s}^\prime$ by 
$L_{\rm s,obs}=\Gamma^4 L_{\rm s}^\prime$.
The intrinsic power emitted over the entire solid angle equals
$\Gamma^2L_{\rm s}^\prime$.
We can then relate the synchrotron power $\Gamma^2 L_{\rm s}^\prime$  
to $L_{\rm k}$ (which is proportional to $n^\prime$)
and $L_{\rm B}$ (which is proportional to the magnetic energy density 
$U_{\rm B}$),
obtaining

\begin{equation}
\Gamma^2 L_{\rm s}^\prime \, =\, 
{16 \sigma_{\rm T} L_{\rm k} L_{\rm B} 
\over 9\pi R m_{\rm e}c^3 \Gamma^2}
\,  { <\gamma^2> \over <\gamma>+m_{\rm p}/m_{\rm e}}
\, =\, 
{2 \sigma_{\rm T} L_{\rm k} B^2 R 
\over 9\pi m_{\rm e}c^2 }
\,  { <\gamma^2> \over <\gamma>+m_{\rm p}/m_{\rm e}}
\end{equation}

Requiring $\Gamma^2 L_{\rm s}^\prime < L_{\rm k}$ implies:

\begin{equation}
{ <\gamma^2> \over <\gamma>+m_{\rm p}/m_{\rm e} } \, <\, 
{9\pi \over 16}\, {R m_{\rm e}c^3 \Gamma^2 \over \sigma_{\rm T}  L_{\rm B}}
\, =\, 
{2\pi \over 16}\, {m_{\rm e}c^2 \over \sigma_{\rm T} R  B^2}
\end{equation}
Assuming a given particle energy distribution $n^\prime (\gamma)$
allows to calculate the left hand side of eq. (5) and therefore
to derive a limit for $\gamma_{\rm peak}$.
For instance, $n^\prime (\gamma)\propto\gamma^{-2}$ between 
$\gamma_{\rm min}< m_{\rm p}/m_{\rm e}$
and $\gamma_{\rm peak}$ yields

\begin{equation}
\gamma_{\rm peak}\, <\, 
{9\pi \over 2 \gamma_{\rm min}}\, {m_{\rm p}c^2  \over \sigma_{\rm T} B^2}\,
=\, 3.2\times 10^6\, {1 \over \gamma_{\rm min} R_{16} B^2}
\end{equation}
Here the notation $Q=10^xQ_x$ is used, with cgs units.
The observed synchrotron peak frequency
$\nu_{\rm peak}=3.7\times 10^6 \gamma_{\rm peak}^2B\Gamma$  Hz corresponds to
\begin{equation}
\nu_{\rm peak} \, =\, 1.6\, {\Gamma_1 \over \gamma_{\rm min}^2 R_{16}^2 B^3} 
\quad {\rm MeV}
\end{equation}

If the synchrotron power $\Gamma^2L_s^\prime$ is of the same order
of $L_{\rm B}$ and $L_{\rm k}$, we obtain the scaling 
of $\gamma_{\rm peak}$ and 
$\nu_{\rm peak}$ with the intrinsic synchrotron power:
$\gamma_{\rm peak}\propto (L_s^\prime)^{-1}$ and
$\nu_{\rm peak}\propto (L_s^\prime)^{-3/2}$.
{\it Only the less powerful sources can have their high energy
peak above the TeV band}.

\begin{figure}
\vskip -2truecm
\psfig{file=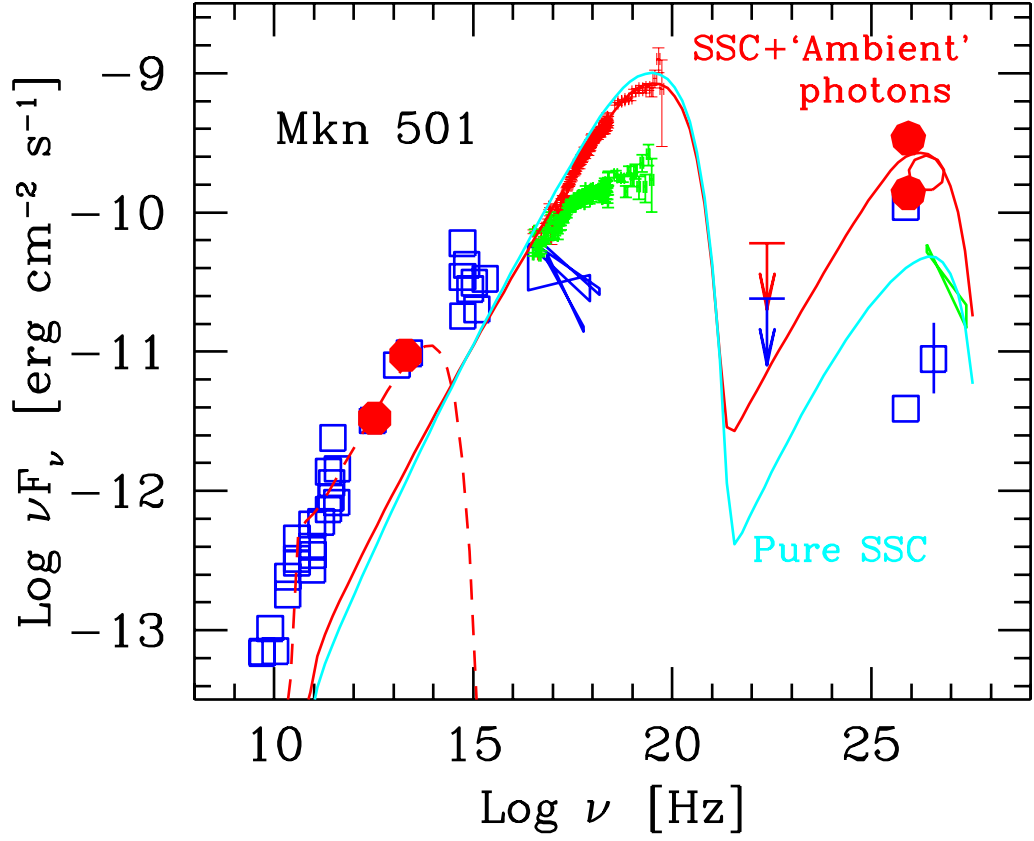} 
\vskip -7 truecm
\caption{The SED of Mkn 501 during the active state of the
1997 flare (filled symbols) is fitted with an homogeneous pure SSC model
and a model in which the IR flux comes from a contiguous region
of the jet, and accounts for the observed mm--IR spectrum (dashed line).}
\end{figure}

\begin{figure}
\vskip -2truecm
\psfig{file=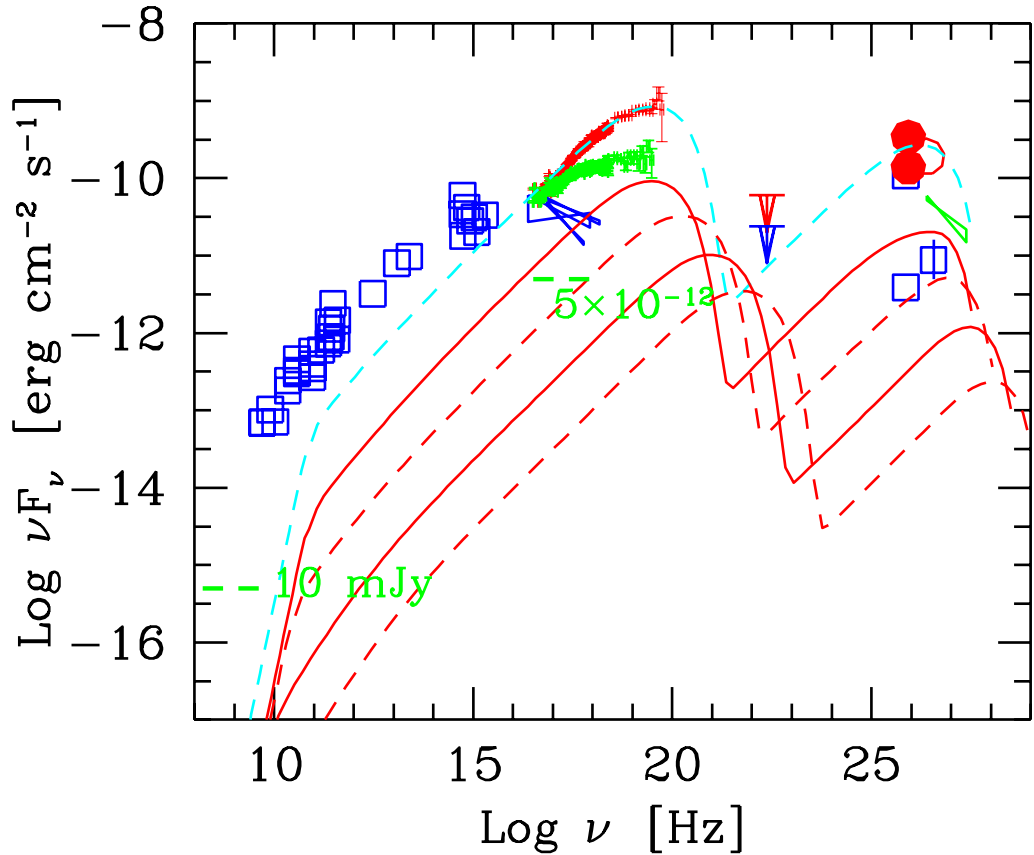} 
\vskip -7 truecm
\caption{SSC and ``ambient" photon models in which the maximum
electron energies increases $\propto 1/L_s^\prime$.
For all models $z=0.1$ is assumed, except the one fitting Mkn 501,
for which $z=0.034$.
The dashed horizontal lines marks the level of 10 mJy at 5 GHz,
approximately the level of the faintest X--ray selected BL Lacs,
and of $5\times 10^{-12}$ erg s$^{-1}$ between 0.3 and 3.5 keV,
approximately the limit flux of the Einstein SLEW survey.  
The SED of Mkn 501 is shown for comparison.}
\end{figure}

\section{The predicted TeV flux of extreme BL Lacs}

Given the scaling just obtained, we can easily
calculate the predicted spectrum of an homogeneous source,
in which the particle distribution is the result of continuous
injection and radiative cooling.
The main uncertainty in this is the amount of soft photons
used as seeds for the Compton scattering process.
Since BL Lacs, especially HBL, have very weak lines and no signs 
of thermal emission (i.e. the blue bump),
it is natural to assume, for these sources, that a pure SSC model applies.
However, it not easy with pure SSC to fit the spectrum of 
Mkn 501 during the 1997 flare and at the same time to account 
for the similar amplitude variability of the synchrotron X--rays 
and the Compton $\gamma$--rays.
This is because the observed flattening of the X--ray spectrum for
brighter states, if extrapolated towards lower frequencies,
constrains the pure SSC model to underpredict the IR flux,
and more severely so for brighter states.
This implies that although the number of high energy electrons increases
during high states, the number of IR photons, which are the seeds
for the Compton process, decreases.
As a result the TeV flux is predicted to vary much less than observed.
This can be ``cured" either by:
\begin{itemize}
\item assuming that the particle distribution {\it is not} the result
of continuous injection and cooling.
For instance, the injection can be impulsive, with the high energy
synchrotron flux strictly following the injection phases, while at lower 
energies, where the cooling time is longer, the particle can accumulate 
and be considered the result of an ``average" (over time) injection.
\item assuming that the active region containing the most energetic
electrons is embedded in a larger region (i.e. a contiguous part 
of the jet) which is much more steady, and that contributes substantially
to the radiation energy density in the IR.
Also this radiation is beamed with the same Doppler factor, since it is
produced by the jet.
\end{itemize}
In both cases the result is that in the active region there is a population
of more steady IR photons contributing to the scattering.
In these conditions the hard X--rays and the TeV flux vary together
with the same amplitude, as observed.
Since this case is somewhat different both from the pure SSC scenario
and from the EC case, I will called it ``ambient" photon model.
The level of the Compton emission is larger than in the pure SSC case,
since we have an extra contribution to the seed photons.
This can be seen in Fig. 5, where I made the comparison between pure
SSC and the ``ambient" photon case, assuming that the extra IR photons
account for the observed IR flux (see also Ghisellini 1998).

Fig. 6 shows some models along these lines.
These are SSC plus ``ambient" photon models in which the maximum
electron energies increases $\propto 1/L_s^\prime$.
Note that in the radio band these models severely underestimate the
observed flux, since the radiaiton is self absorbed at high radio frequencies
in these compact regions. 
Additional components must be present in the source, to account for
the radio emission. 
We can, very roughly, estimate the level of the flux in the radio band 
in the following way:
i) extrapolate the thin synchrotron power law of the model fitting Mkn 501 
towards low frequencies, obtaining a factor 10 of discrepancy between 
the extrapolation and the data;
ii) do the same extrapolation with the other curves, and multiply
the extrapolated flux by the same factor 10.

\section{Where are very extreme BL Lacs?}

Fig. 6 reports for illustration the level of radio flux of 10 mJy at 5 GHz, 
appropriate for the large area deep radio survey, and the flux of 
$5\times 10^{-12}$ erg cm$^{-2}$ s$^{-1}$ in X--rays, which is approximately 
the level of the SLEW survey (Perlman et al. 1996).
It can be seen that objects like Mkn 501 could have been detected
in existing large area X--ray survey (i.e. the ROSAT RASS), even if
they lye at a redshift of 0.1 (i.e. 3 times more distant than Mkn 501).
In this case the host galaxy contributes in the optical band as in Mkn 501,
and these objects can be easily classified as BL Lacs.
In the RASS (and possibly SLEW) survey there can therefore exist
extreme BL Lacs. 
However, for the majority of these already detected and classified objects
we only know the radio, optical and X--ray fluxes.
It is therefore interesting to find a tool able to recognize extreme
BL Lacs only on the basis of these fluxes and the corresponding
broad band spectral indices, as done e.g. by Fossati (1998).
Using this this tool we have selected a handful of BL Lac objects
to be observed with BeppoSAX
(0120+340, 0224+014, 0120+340, 0548--322, 1101--232, 1320+084, 1426+428, 
2005--489, 2356-309): 
if their X--ray spectral index will turn
out to be flat ($\alpha_x< 1$), this would guarantee that their
synchrotron spectrum peaks at high energies, and hence they would
be good candidates for detection in the TeV band.

Decreasing the intrinsic power still further, we may find objects
whose synchrotron peak is at energies even higher.
In the soft X--ray range, the reduced flux could let these objects
escape detection in large area X--ray surveys.
In the radio, the flux can be below the $\sim$10 mJy level.
Furthermore, even if detected in X--ray surveys, it may be difficult to 
classify these objects as BL Lacs, because in the optical the contrast 
between the galaxy and the non--thermal emission is reduced.
It may be that the easiest way to find them is through MeV and TeV surveys. 
VERITAS can therefore play a crucial role in this respect, finding 
conspicuous TeV sources that look like normal elliptical galaxies 
in other bands.

How many of these sources do we expect?
Since they are intrinsically weak BL Lacs, one expects to
have a lot of them.
The exact answer however depends on the lower end of the BL Lac
luminosity function (which is still unknown), where these extreme 
sources are predicted to be.
One can find a solid upper limit to the number of these
sources by requiring that their MeV emission does not overproduce
the MeV $\gamma$--ray background (Ghisellini et al., in preparation).

\section{Conclusions}
The blazar sequence discussed in this paper suggests a key role
for future, more sensitive Cherenkov detectors, such as VERITAS.
The most extreme (i.e. emitting at the higher energies) sources
should in fact be the intrinsically weakest, therefore the 
most numerous.
There is the possibility that a new subclass of BL Lacs exists,
whose synchrotron spectrum peaks in the MeV band and
whose Compton spectrum peaks at or even above one TeV, and whose
existence may be discovered by instruments like VERITAS, in the TeV
band, or like INTEGRAL, in the MeV band.

Very extreme BL Lacs may be the sources where the dissipation of the
power carried by the jet is the most efficient, and where we
can study the acceleration mechanism at its limit.
If they exist, and if they can be detected, they will be very useful
for the determination of the IR background: since their peak is above
one TeV, it will be much less ambiguous to disentangle the
effect of photon--photon absorption from the intrinsic curvature
of the spectrum.

\vskip 1 true cm
{\bf Acknowledgments}

It is a pleasure to thank Annalisa Celotti and Laura Maraschi for 
constant help and for years of fruitful discussions

\end{document}